\documentclass[12pt]{article}
\usepackage{epsf}
\usepackage{graphicx}
\def\hybrid{\topmargin 0pt      \oddsidemargin 0pt
	\headheight 0pt \headsep 0pt
	\textheight 9in         
	\textwidth 6.25in       
	\marginparwidth .875in
	\parskip 5pt plus 1pt   \jot = 1.5ex}

\catcode`\@=11
\def\marginnote#1{}
\newcount\hour
\newcount\minute
\newtoks\amorpm
\hour=\time\divide\hour by60
\minute=\time{\multiply\hour by60 \global\advance\minute by-\hour}
\edef\standardtime{{\ifnum\hour<12 \global\amorpm={am}%
	\else\global\amorpm={pm}\advance\hour by-12 \fi
	\ifnum\hour=0 \hour=12 \fi
	\number\hour:\ifnum\minute<10 0\fi\number\minute\the\amorpm}}
\edef\militarytime{\number\hour:\ifnum\minute<10 0\fi\number\minute}

\def\draftlabel#1{{\@bsphack\if@filesw {\let\thepage\relax
   \xdef\@gtempa{\write\@auxout{\string
      \newlabel{#1}{{\@currentlabel}{\thepage}}}}}\@gtempa
   \if@nobreak \ifvmode\nobreak\fi\fi\fi\@esphack}
	\gdef\@eqnlabel{#1}}
\def\@eqnlabel{}
\def\@vacuum{}
\def\draftmarginnote#1{\marginpar{\raggedright\scriptsize\tt#1}}

\def\draft{\oddsidemargin -.5truein
	\def\@oddfoot{\sl preliminary draft \hfil
	\rm\thepage\hfil\sl\today\quad\militarytime}
	\let\@evenfoot\@oddfoot \overfullrule 3pt
	\let\label=\draftlabel
\let\marginnote=\draftmarginnote
	\let\marginnote=\draftmarginnote
   \def\@eqnnum{(\theequation)\rlap{\kern\marginparsep\tt\@eqnlabel}%
\global\let\@eqnlabel\@vacuum}  }

\def\numberbysection{\@addtoreset{equation}{section}
	\def\theequation{\thesection.\arabic{equation}}}

\def\underline#1{\relax\ifmmode\@@underline#1\else
	$\@@underline{\hbox{#1}}$\relax\fi}

\def\titlepage{\@restonecolfalse\if@twocolumn\@restonecoltrue\onecolumn
     \else \newpage \fi \thispagestyle{empty}\c@page\z@
	\def\thefootnote{\fnsymbol{footnote}} }

\def\endtitlepage{\if@restonecol\twocolumn \else  \fi
	\def\thefootnote{\arabic{footnote}}
	\setcounter{footnote}{0}}  
\catcode`@=12
\relax

\def\beq{\begin{equation}}
\def\eeq{\end{equation}}
\def\bea{\begin{eqnarray}}
\def\eea{\end{eqnarray}}
\def\nn{\nonumber}

\relax
\hybrid

\begin{document}

\begin{titlepage}
\begin{center}
October~2006 \hfill . \\[.5in]
{\large\bf Renormalization group flows for the second $Z_{5}$ parafermionic field theory.} 
\\[.5in] 
{\bf Vladimir S.~Dotsenko 
 and Benoit Estienne}\\[.2in]
 {\it Laboratoire de Physique Th\'eorique 
et Hautes Energies}\footnote{Unit\'e Mixte de Recherche UMR 7589}\\

 {\it Universit{\'e} Pierre et Marie Curie, Paris-6; CNRS\\
              Universit{\'e} Denis Diderot, Paris-7\\
               Bo\^{\i}te 126, Tour 25, 5\`eme {\'e}tage,
               4 place Jussieu, F-75252 Paris Cedex 05, France.}\\

\end{center}

\underline{Abstract.}

Using the renormalization group approach, the Coulomb gas and the coset techniques, 
the effect of slightly relevant perturbations is studied 
for the second parafermionic field theory with the symmetry $Z_{5}$. 
New fixed points are found and classified.

\end{titlepage}

\newpage

While the first series of parafermionic conformal field theories [1] 
is well studied and applied in various domains [2,3,4], 
the second parafermionic series, with the symmetry $Z_{N}$, 
has been developed fairly recently [5-8] and it still awaits its applications.

In the case of the first series, to a given $N$ (of $Z_{N}$) 
is associated a single conformal theory. 
This is different for the second series: for a given $Z_{N}$, 
there exist an infinity of unitary conformal theories $Z_{N,p}$, 
with $p=N-2+k$, $k=1,2,3,...\infty$. These theories correspond 
to degenerate representations of the corresponding parafermionic chiral algebra. 
They are much more rich in their content of physical fields, 
as compared to the theories of the first series. 
They are also much more complicated. But, on the other hand, 
the presence of the parameter $p$, for a given $Z_{N}$, 
opens a way to reliable perturbative studies. It allows in particular 
to study the renormalisation group flows 
in the space of these conformal theory models,
under various perturbations.

 In this paper we shall present results for a particular case of this problem: for the renormalization group flows of the $Z_{5,p}$ theories, being perturbed by two slightly relevant fields.

The details of the $Z_{5}$ parafermionic theory, the second one, could be found in [5]. The $q$ charge of $Z_{5}$ takes values
$q=0,\pm 1,\pm 2$, so that in the Kac table of this theory one finds the $Z_{5}$ neutral fields, of $q=0$, the $q=\pm 1$ and the $q=\pm 2$ doublets, and the $Z_{2}$ disorder fields. The symmetry of the theory is actually $D_{5}$, which is made of $Z_{5}$ rotations and the $Z_{2}$ reflections in 5 different axises. These last symmetry elements amount to the charge conjugation symmetry: $q\rightarrow -q$.

 We want to perturbe by the $Z_{5}$ ($D_{5}$ in fact) neutral fields, in order to preserve this symmetry. Perturbatively well controlled domain of $Z_{5,p}$ theories is that of $ p\gg 1$, giving a small parameter $\epsilon\sim 1/p$. This is similar to the original perturbative renormalization group treatment of minimal models for Virasoro algebra based conformal theory [9,10].

In this domain, i.e. for $p\gg 1$, one finds, in the lower part of the Kac table of $Z_{5}$ parafermionic theory, two $Z_{5}$ neutral fields which are slightly relevant 
and which close by the operator algebra. They are:
\beq
S=\Phi_{(1,1\mid 31)}
\eeq
\beq
A=A_{-\frac{2}{5}}\Phi_{(1,1\mid 1,3)}
\eeq
The first one is a $Z_{5}$ singlet and the second is a parafermionic algebra descendant of a doublet $q=1$ field. They both belong to the neutral $q=0$ sector of $Z_{5}$ and they are both  Virasoro algebra primaries.

Their labeling as $S$ and $A$ is just our shortened notations (in this paper) for these fields.

In general, the parafermionic algebra primaries of the second $Z_{5}$ conformal theory are labeled by double indices of two ($\alpha_{+}$ and $\alpha_{-}$) lattices of the $B_{2}$ classical Lie algebra [5]:
\beq
\Phi_{(n_{1},n_{2}\mid n'_{1},n'_{2})}
\eeq
The first and second couples of indices correspond respectively to the  $\alpha_{+}$ and $\alpha_{-}$ $B_{2}$ lattices.
$\alpha_{+}$ and $\alpha_{-}$ are the usual Coulomb gas type parameters. For the $Z_{5,p}$ conformal theory they take the values:
\beq
\alpha_{+}=\sqrt{\frac{p+2}{p}},\quad\alpha_{-}=-\sqrt{\frac{p}{p+2}}
\eeq

The formulas for the conformal dimensions of the fields (3) could be found in [5]. One could check that the dimensions of the fields $S$ and $A$ in (1) and (2) have the following values:
\beq
\Delta_{S}=\frac{5}{2}\alpha^{2}_{-}-\frac{3}{2}=1-5\epsilon
\eeq
\beq
\Delta_{A}=\frac{3}{2}\alpha_{-}^{2}-\frac{1}{2}=1-3\epsilon
\eeq
We have defined $\epsilon$ as follows:
\beq
\alpha^{2}_{-}=\frac{p}{p+2}=1-\frac{2}{p+2}=1-2\epsilon,\quad\quad
\epsilon=\frac{1}{p+2}\simeq\frac{1}{p}
\eeq

Perturbing with the fields $S$ and $A$ corresponds to taking the action of the theory in the form:
\beq
{\bf A}={\bf A}_{0}+\frac{2g}{\pi}\int d^{2}xS(x)+\frac{2h}{\pi}\int d^{2}x A(x) 
\eeq
where $g$ and $h$ are the corresponding coupling constants; 
the additional factors  $\frac{2}{\pi}$  are added to simplify the coefficients 
of the renormalization group equations which follow;  ${\bf A}_{0}$ is assumed 
to be the action of the unperturbed $Z_{5,p}$ conformal theory.

 It will be shown below that the operator algebra of the fields $S$ and $A$ is of the form:
\beq
S(x')S(x)=\frac{D_{1}}{|x'-x|^{4\Delta_{S}-2\Delta_{A}}}A(x)+...
\eeq
\beq
A(x')A(x)=\frac{D_{2}}{|x'-x|^{2\Delta_{A}}} A(x)+...
\eeq
\beq
S(x')A(x)=\frac{D_{1}}{|x'-x|^{2\Delta_{A}}} S(x)+...
\eeq
Only the fields which are relevant for the renormalization group flows are shown explicitly in the r.h.s. of the equations (9)-(11). For instance, the identity operator is not shown in the r.h.s. of (9) and (10) while it is naturally present there.
The operator algebra constants in (9) and (11) should 
obviously be equal, as the two equations could be related to 
a single correlation function     $<S(x_{1})S(x_{2})A(x_{3})>$.

Assuming the operator algebra expansions in (9)-(11), one finds, in a standard way, the following renormalization group equations for the couplings $g$ and $h$:
\beq
\frac{dg}{d\xi}=2\cdot 5\epsilon\cdot g-4D_{1}gh
\eeq
\beq
\frac{dh}{d\xi}=2\cdot 3\epsilon\cdot h-2D_{2}h^{2}-2D_{1}g^{2}
\eeq
These are up to (including) the first non-trivial order of the perturbations in $g$ and $h$.

 The problem now amounts to justifying the operator algebra expansions in (9) - (11) and to calculating the constants $D_{1}$ and $D_{2}$.

The efficient method for calculating the operator product expansions and defining the corresponding coefficients, is that of the Coulomb gas technique.

 Calculating directly the expansions of the products of the operators (1), (2) encounters a problem: the explicit form of the Coulomb gas representation for the $Z_{5}$ theory is not known. We shall get around this problem by using the coset representation for the (second) $Z_{5}$ theory and the related techniques. In particular, we shall generalize the methode developed in papers [11,12] for the $SU(2)$ cosets.

The second $Z_{5}$ theory, which we shall denote as $Z_{5}^{(2)}$, could be represented as the following coset construction [13]:
\beq
Z^{(2)}_{5,p}=\frac{SO_{k}(5)\times SO_{2}(5)}{SO_{k+2}(5)}
\eeq
Here $SO_{k}(5)$ is the orthogonal affine algebra of level $k$; $p=3+k$. This coset could be rewritten as follows:
\beq
Z^{(2)}_{5,p}\times\frac{SO_{1}(5)\times SO_{1}(5)}{SO_{2}(5)}=\frac{SO_{k}(5)\times SO_{1}(5)}{SO_{k+1}(5)}\times\frac{SO_{k+1}(5)\times SO_{1}(5)}{SO_{k+2}(5)}
\eeq
The two coset factors in the r.h.s., as well as the additional coset factor in the l.h.s., correspond to the $WB_{2}$ theories [14]. For these theories the Coulomb gas representation is known. It is made of two bosonic fields, quantized with a background charge and the Ising model fields: $\Psi$ (free fermion) and $\sigma$ (spin operator) [14].

 The equation (15) could be rewritten as
\beq
Z^{(2)}_{5,p}\times WB_{2,1}=WB_{2,k}\times WB_{2,k+1}
\eeq
This equation relates the representations of the corresponding algebras. It could be
reexpressed in terms of characters of representations, as is being usually done in the analyses of cosets. But this equation allows also to relate the conformal blocs of correlation functions. In doing so one relates the chiral (holomorphic) factors of physical operators. This later approach has been developped and analyzed in great detail in the papers [11,12], for the $SU(2)$ coset theories.

As it was said above, the chiral factor operators are related to the conformal bloc functions, not to the actual physical correlators. On the other hand, the coefficients of the operator algebra expansions are defined by the three point functions. These latter are factorizable, into holomorphic - antiholomorphic functions. So that, when the relation is established on the level of chiral factor operators, for the holomorphic three point functions, this relation could then be easily lifted to the relation for the physical correlation functions. Saying it differently, with the relations for the chiral factor operators one should be able to define the square roots of the physical operator algebra constants.

By matching the conformal dimensions of operators on the two sides of the coset equation (16) one finds the following decompositions for the operators $S$ and $A$ in (1) and (2) (their chiral factors in fact):
\beq
\Phi_{(11\mid 31)}^{(Z_{5,p})}\times \Phi_{(11\mid 11)}^{(WB_{2,1})}=\Phi_{(11\mid 21)}^{(WB_{2,k})}\times\Phi_{(21\mid 31)}^{(WB_{2,k+1})}
\eeq
\beq
A_{-\frac{2}{5}}\Phi_{(1,1\mid 1,3)}^{(Z_{5,p})}\times 
\Phi_{(11\mid 11)}^{(WB_{2,1})}
=a\,\,\Phi_{(11\mid 11)}^{(WB_{2,k})}\times\Phi_{(11\mid 13)}^{(WB_{2,k+1})}
+b\,\,\Phi_{(11\mid 13)}^{(WB_{2,k})}\times\Phi_{(13\mid 33)}^{(WB_{2,k+1})}
\eeq
The coefficients $a$ and $b$ in (18) are still to be determined. 
$\Phi^{(...)}_{(11\mid 11)}$ are the identity operators. They could actually be dropped. But we shall keep them sometimes, when this makes the decomposition more explicit. The operator $\Phi_{(11\mid 11)}^{(WB_{2,1})}$ in the l.h.s. of (17), (18)
could be definitely suppressed.

By equations (17), (18), one observes that to decompose the products $SS$, $AA$, $SA$, as in eqs.(9)-(11), one needs to know the decompositions of products of the operators of $WB_{2,k}$ and $WB_{2,k+1}$ theories: 
$\Phi_{(11\mid 21)}^{(WB_{2,k})}\times\Phi^{(WB_{2,k})}_{(11\mid 21)}$, $\Phi_{(21\mid 31)}^{(WB_{2,k+1})}\times\Phi_{(21\mid 31)}^{(WB_{2,k+1})}$, etc.

These could be defined by using the Coulomb gas representation of the $WB_{2}$ theory. 
We shall give below the results of our analyses. 
The details of the calculations as well as more detailed analyses 
will be given in our next paper [15]. 

We have found the following values for the operator algebra constants of $WB_{2}$:
\bea
D_{(11\mid 13)(11\mid 13)(11\mid 13)}=\frac{3\sqrt{2}}{\sqrt{5}},\nn\\
D_{(11\mid 13)(13\mid 13)(13\mid 13)}=\frac{3\sqrt{2}}{\sqrt{5}}\epsilon^{2},\nn\\
D_{(11\mid 21)(11\mid 21)(11\mid 13)}=\frac{2\sqrt{2}}{\sqrt{5}},\nn\\
D_{(21\mid 31)(21\mid 31)(11\mid 13)}=\frac{\sqrt{5}}{2\sqrt{2}},\nn\\
D_{(13\mid 13)(13\mid 13)(13\mid 13)}=1,\nn\\
D_{(21\mid 31)(21\mid 31)(13\mid 13)}=\frac{5}{8}
\eea
These are the constants which are needed for our calculations. 
We give their values to the leading order in $\epsilon$, 
which is sufficient for the renormalization group equations in (12), (13).

Using the above values of the $WB_{2}$ constants and performing the expansions for the 
products of operators in (17), (18), one finds that the products $SS$, $AA$, $SA$ have the expansions of the form given in eqs.(9)-(11), with the constants $D_{1}$, $D_{2}$
having the following values:
\beq
D_{1}=\sqrt{\frac{5}{2}},\quad D_{2}=\frac{3}{\sqrt{10}}
\eeq
In the process of these calculations one defines also the coefficients $a$ and $b$ in (18): $a=b=1/\sqrt{2}$.

Some remarks are in order. 

Doing the expansions of the products of $WB_{2,k}$ and $WB_{2,k+1}$ operators 
one does them:

1) with the square roots of the constants in (19);

2) one keeps in these expansions the "diagonal" cross-products only: the products of $WB_{2,k}$ and $WB_{2,k+1}$ operators which appear in coset relations for operators, 
due to eq.(16).

For operators, the relation (16) implies the decompositions of the following general form:
\bea
\Phi^{(Z^{(2)}_{5,p})}_{(n_{1},n_{2}\mid n'_{1},n'_{2})}\times\Phi^{(WB_{2,1})}_{(s_{1},s_{2}\mid s'_{1},s'_{2})}\nn\\
=\sum_{l_{1},l_{2}} a(l_{1},l_{2}) \Phi^{(WB_{2,k})}_{(n_{1}n_{2}\mid l_{1},l_{2})}\times
\Phi^{(WB_{2,k+1})}_{(l_{1},l_{2}\mid n'_{1},n'_{2})}
\eea
The operators in this relation could be primaries or their descendants. The equations (17), (18) are two particular examples of eq.(21).
These decompositions will be discussed in more detail in [15]. They generalize the corresponding relations for the  $SU(2)$ cosets of [11,12].

The "diagonal" cross-products correspond to products (of $WB_{2,k}$ and $WB_{2,k+1}$ operators) of the type which appear in the r.h.s. of (21). The rest of possible cross-products have to be dropped when doing expansions.

The above features, square roots of constants and keeping the diagonal terms only, are due to the fact that we are dealing with the conformal bloc functions and not with the actual physical correlators. These features are discussed in much detail in the paper [12].

Equally, the overal factors of the resulting expressions (the expressions which should correspond to the decomposed $Z_{5}^{(2)}$ operators) provide the square roots of the $Z_{5}^{(2)}$ structure constants, and not the constants themselves. In this way one obtains the square roots of the values of $D_{1}$, $D_{2}$ in eq.(20).

Substituting now the values of $D_{1}$, $D_{2}$ into the renormalization group equations (12), (13) and analysing them by the standard methods one obtains the following results.

The phase diagram of constants $g$ and $h$ contains:

1) the initial fixed point $g_{0}^{\ast}=h^{\ast}_{0}=0$;

2) the fixed point on the $h$ axis: $g^{\ast}_{1}=0$, $h_{1}^{\ast}=\sqrt{10}\epsilon$;

3) two additional fixed points for non-vanishing values of the two couplings: 
$g^{\ast}_{2}=\sqrt{\frac{3}{2}}\epsilon$, $h^{\ast}_{2}=\frac{\sqrt{10}}{2}\epsilon$, 
and $g^{\ast}_{3}=-\sqrt{\frac{3}{2}}\epsilon$, $h^{\ast}_{3}=\frac{\sqrt{10}}{2}\epsilon$.

\begin{centering}
\includegraphics[scale=0.8]{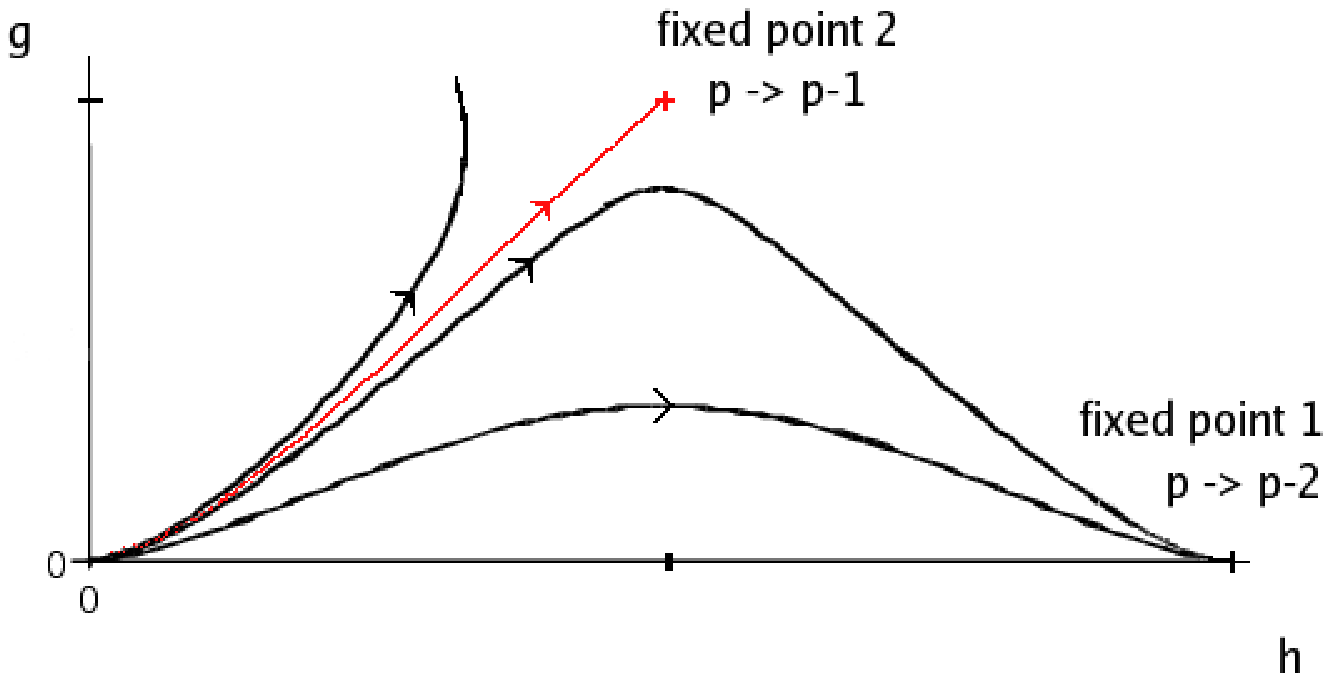}
\end{centering}

\noindent The renormalization group flows are shown in the figure. 
They are symmetrical with respect to $g \rightarrow -g$.

The value of the central charge at the point $g^{\ast}_{1}=0$, 
$h^{\ast}_{1}=\sqrt{10}\epsilon$ agrees with that 
of the theory $Z^{(2)}_{5,p-2}$. This confirms the observation, 
made with the $SU(2)$ cosets [11,12] and, more generally, with the cosets 
for the simply laced algebras [16], that the perturbation of a coset theory
caused by an appropriate operator drives $p$ to $p-\Delta p$, 
$\Delta p$ being equal to the shift parameter of the coset.
In our case the shift parameter of the coset is equal to 2, eq.(14). 
Note that the algebra $B_{2}\equiv SO(5)$ is not a simply laced one.

On the other hand, the appearance of two extra fixed points, 
$(g^{\ast}_{2}$, $h^{\ast}_{2})$ 
and ($g^{\ast}_{3}$, $h^{\ast}_{3}$), is somewhat surprising. By the value
of the central charge, the two critical points 
correspond to the theory $Z^{(2)}_{5,p-1}$. This assignment has further been
verified by calculating the critical dimension 
of the operator $\Phi_{(1,n\mid1,n)}$ at these points.

We observe that such additional fixed points do not appear 
in the parafermionic model $Z_{3}^{(2)}$: 
the second $Z_{3}$ parafermionic theory 
with $\Delta_{\Psi}=4/3$ [17]. This model could be realized 
by the $SU(2)$ cosets. Its perturbations, 
with two slightly relevant operators, have been analysed in [11,12].

Further analysis and discussions will be left for the paper [15].

{\bf Acknoledgements:} Usuful discussion with V.A.Fateev 
is gratefully acknowledged.

\end{document}